\documentclass[prx,aps,twocolumn,superscriptaddress]{revtex4-1}
\usepackage{graphicx}
\usepackage{rotating}
\usepackage{amsmath}
\usepackage{latexsym}
\usepackage{color}

\begin{document}

\title{Quantifying the overall characteristics of urban mobility considering spatial information}

\author{Hao Wang}
\affiliation{School of Urban Planning and Design, Shenzhen Graduate School, Peking University, Shenzhen 518055, China}

\author{Pengjun Zhao} \email{pengjun.zhao@pku.edu.cn}
\affiliation{School of Urban Planning and Design, Shenzhen Graduate School, Peking University, Shenzhen 518055, China}
\affiliation{School of Urban and Environmental Sciences, Peking University, Beijing 100871, China}

\author{Xiao-Yong Yan} \email{yanxy@bjtu.edu.cn}
\affiliation{School of Systems Science, Beijing Jiaotong University, Beijing 100044, China}

\begin{abstract}
Quantification of the overall characteristics of urban mobility using coarse-grained methods is crucial for urban management, planning and sustainable development. Although some recent studies have provided quantification methods for coarse-grained numerical information regarding urban mobility, a method that can simultaneously capture numerical and spatial information remains an outstanding problem. Here, we use mathematical vectors to depict human mobility, with mobility magnitude representing numerical information and mobility direction representing spatial information. We then define anisotropy and centripetality metrics by vector computation to measure imbalance in direction distribution and orientation toward the city center of mobility flows, respectively. As a case study, we apply our method to 60 Chinese cities and identify three mobility patterns: strong monocentric, weak monocentric and polycentric. To better understand mobility pattern, we further study the allometric scaling of the average commuting distance and the spatiotemporal variations of the two metrics in different patterns. Finally, we build a microscopic model to explain the key mechanisms driving the diversity in anisotropy and centripetality. Our work offers a comprehensive method that considers both numerical and spatial information to quantify and classify the overall characteristics of urban mobility, enhancing our understanding of the structure and evolution of urban mobility systems.

\end{abstract}
\maketitle

\section{Introduction} \label{sec:intro}
Our world is experiencing rapid urbanization, and nearly $70\%$ of the world’s population is projected to be living in cities by 2050, making it imperative to understand the phenomena observed in urban systems~\cite{Barthelemy19,Yabe22,Mc16,Brelsford17}. The quantitative characterization of human mobility is at the core of urban studies~\cite{Brockmann06,Gonz08,Song10,Simini12,Yan17,Gallotti16,Barbosa18,Aless20,Schlapfer21,Li22,Lu12}, as it is relevant to many practical applications ranging from epidemic containment~\cite{Jia20,Chang21,Vahedi21} to traffic prediction~\cite{Dios11,Yan14} and urban planning~\cite{Batty08}. A central issue in this field is to identify coarse-grained metrics that can describe the overall characteristics of urban mobility on large scales~\cite{Louail15,Louail14,Lee17,Bassolas19,Golden99}. The coarse-grained metrics are simple enough to grasp the main characteristics but complex enough to capture the essential aspects of urban mobility systems, contributing to the evaluation of urban mobility systems in the context of the worldwide impetus to create liveable cities~\cite{Rubenstein17,Hirsch05,Hoel13,Wieder16,Bagrow12}. For instance, governments can objectively compare the mobility characteristics across cities according to these metrics and implement relevant policies to guide urban transformations toward sustainability and health. Therefore, these metrics will undoubtedly be of considerable help in achieving better management and planning of complex urban systems, which is highly important for the sustainable development goals set out by the United Nations~\cite{Lu15,Weiss18,Li20}.

In the past decade there has been a great advance in coarse-grained metrics for the overall characteristics of urban mobility. Louail \emph{et al}.~\cite{Louail15} proposed the hotspot flow matrix, and later Bassolas \emph{et al}.~\cite{Bassolas19} extended it and developed the flow-hierarchy to extract coarse-grained numerical information from an origin-destination (OD) matrix. Although these metrics have become remarkably successful in quantifying the hierarchical structural characteristics of urban mobility, they do not take into account the spatial information associated with urban mobility, such as the position of the location or the direction of mobility~\cite{Wieder16,Daqing11,Barthelemy22}. Spatial information is a fundamental aspect of mobility since human mobility typically occurs in spatial domains~\cite{Barthelemy11}. For example, one mobility network can have multiple spatial layouts even if they have the same OD matrix (see  Supplementary Note 1 and  Supplementary Fig. 1). Thus, it is insufficient to fully describe the overall characteristics of urban mobility without considering spatial information~\cite{Barthelemy11,Liu21}.

Recently, some studies have encoded the spatial information of urban mobility when investigating urban structure and dynamics. Examples include the urban dilatation metric, which can measure the evolution of the average distance between individuals during the day~\cite{Louail14}, the route geometric metric, which can capture the tendency of travel routes to gravitate toward a city center~\cite{Lee17}, the mean velocity vector, which chooses the dominant velocity direction as the discretized representative direction of the location~\cite{Shida20,Shida22}, and the wayfinding performance metric, which can quantify the spatial navigation ability of humans~\cite{Coutrot22}. However, these metrics do not take into account the mobility flow between locations; that is, the numerical information of mobility is ignored. Excitingly, Mazzoli \emph{et al}. proposed a field theory for recurrent mobility~\cite{Mazzoli19} that incorporates both numerical and spatial information regarding urban mobility in the form of vectors and further defined the potential for the vector field that enable intuitive recognition of the overall characteristics of urban mobility at the mesoscopic scale. Despite the great success of the field theory for recurrent mobility, this theory does not produce quantification metrics of the overall characteristics of urban mobility. Therefore, the development of a method that can simultaneously capture numerical and spatial information to create metrics for quantifying the overall characteristics of urban mobility remains an outstanding problem.

Here, we present a method that incorporates both the magnitude and direction of mobility flows to quantify the overall characteristics associated with urban mobility for 60 Chinese cities. We first propose a population mobility vector and introduce two metrics named anisotropy and centripetality that capture the imbalance in the direction distribution and orientation toward the city center of mobility flows, respectively. The two metrics enable us to classify cities and analyze the corresponding urban mobility patterns. Next, we study the relationship between the average commuting distance and city size across different mobility patterns. Then, we present a detailed analysis of the spatiotemporal variations in the two metrics, deepening our understanding of human mobility patterns. Finally, we build a microscopic model to provide a mechanistic explanation for the diverse overall characteristics of urban mobility. Our work provides an alternative tool for quantifying and classifying the overall characteristics associated with urban mobility, as well as analyzing their spatiotemporal variations, which can provide support for urban and transportation planning.

\section{Results}

\subsection{Anisotropy and centripetality of mobility flows}

The OD matrix is essential for quantifying the overall characteristics of urban mobility since it contains detailed information about the flow of people at a given spatial scale and period [11].
Specifically, an OD matrix is an $m \times n$ matrix, where $m$ is the number of origins, $n$ is the number of destinations, $T_{ij}$ is the flow from origin $i$ $(i=1, 2, \dots , m)$ to destination $j$ $(j=1, 2, \dots , n)$, and $O_{i}$ is the outflow from location $i$, with $O_i=\sum_{j\neq i} T_{ij}$.
The OD matrix utilized in this research is obtained from a mobile phone dataset collected over a 2-month period in 60 major Chinese cities (Methods). To explore both the numerical and spatial components of population flows, we partition each urban area (see Methods, Supplementary Note 2 and Supplementary Fig. 2 for details) into a high-resolution square grid and extract the OD matrix and the coordinates of each grid. We first extract the numerical and spatial information associated with human mobility at the location level. Typically, the mobility flow $T_{ij}$ represents the number of individuals who start a journey from $i$ to $j$, where $i$ and $j$ have coordinates. Hence, $T_{ij}$ has both magnitude and direction, and it is appropriate to use mathematical vectors to describe this quantity.
To incorporate the direction associated with human mobility, we vectorize the flow $T_{ij}$ and obtain a vector $T_{ij}\overrightarrow{u_{ij}}$, where $\overrightarrow{u_{ij}}$ is the unit vector from origin $i$ to destination $j$. Next, to factor out the effects of varying outflow across locations~\cite{Lee17,Bassolas19,Mazzoli19}, we normalize the vector $T_{ij}\overrightarrow{u_{ij}}$ by the outflow of its origin, yielding $\overrightarrow{T_{ij}} = T_{ij}\overrightarrow{u_{ij}} / O_i$ (Fig.~\ref{fig1}a). Finally, we sum the vectors pointing to all destinations $j$ (Fig.~\ref{fig1}b) and define this vector sum as the population mobility vector (PMV)
\begin{equation} \label{eq1}
	\overrightarrow{T_{i}} = \sum_{j\neq i} \frac{T_{ij}}{O_i} \overrightarrow{u_{ij}}.
\end{equation}
The PMV reflects the dominant direction among all directions~\cite{Zhang20}, which is a trade-off among $i$’s $\overrightarrow{T_{ij}}$ in different directions.

The PMV is a vector that includes magnitude $\lambda_i$ and direction $\theta_i$. For a given location $i$, the magnitude of the PMV is defined as
\begin{equation} \label{eq2}
	\lambda_i = \mid \overrightarrow{T_{i}} \mid = \mid \sum_{j\neq i} \frac{T_{ij}}{O_i} \overrightarrow{u_{ij}} \mid,
\end{equation}
where $\lambda_i$ ranges from 0 to 1. It can be seen that $\lambda_i$ is 1 if all flows $T_{ij}$ from location $i$ have the same orientation and 0 if the flows $T_{ij}$ from location $i$ are spatially symmetric.
Therefore, the direction-dependent property $\lambda_i$ indicates the imbalance degree in the direction distribution of outflow, and we refer to $\lambda_i$ as the \emph{anisotropy} of outflow from location $i$. The higher the value of $\lambda_i$ (the more concentrated the flows are in a specific direction), the stronger the anisotropy is, and vice versa. After calculating the anisotropy of each location, we compute the anisotropy of urban mobility by taking a weighted average of each location’s anisotropy, with the outflow of each location as the weight~\cite{Coutrot22}
\begin{equation} \label{eq3}
	{\mit \Lambda} = \frac{\sum_i O_i \lambda_i}{\sum_i O_i},
\end{equation}
where $\mit \Lambda$ is the anisotropy of urban mobility, reflecting the overall imbalance degree in the direction distribution of all flows.

For the direction $\theta_i$, we use a common reference point to consistently measure $\theta_i$ across locations. Specifically, we define $\theta_i$ as the relative angle to the city center $C$ (see Methods, Supplementary Note 3 and Supplementary Fig. 3 for the identification of the city center), which is also called the allocentric direction~\cite{Ormond22}, that is, the mobility angle in reference to the city center (see Fig.~\ref{fig1}c). We then calculate the direction $\theta_i$ of the PMV using the law of cosines
\begin{equation} \label{eq4}
	\theta_i = \frac{\overrightarrow{T_{i}} \cdot \overrightarrow{u_{iC}}}
	{\mid \overrightarrow{T_{i}} \mid \mid \overrightarrow{u_{iC}} \mid},
\end{equation}
where $\overrightarrow{u_{iC}}$ is the unit vector from $i$ to $C$ and $\theta_i$ ranges from 0 to $\pi$. It can be seen that $\theta_i$ is 0 if $\overrightarrow{T_{i}}$ is oriented toward the city center and $\pi$ if $\overrightarrow{T_{i}}$ is oriented in the opposite direction. Therefore, the direction $\theta_i$ indicates the orientation degree toward the city center of the dominant mobility, and we can define the \emph{centripetality} of outflow from location $i$ as
\begin{equation} \label{eq5}
	\gamma_i = 1 - \frac{\theta_i}{\pi},
\end{equation}
where $\gamma_i$ ranges from 0 to 1. The higher the value of $\gamma_i$ (the more oriented toward the city center the dominant mobility), the stronger the centripetality is, and vice versa. Similarly, after calculating the centripetality of each location, we compute the centripetality of urban mobility by taking the weighted average of each location's centripetality, with the outflow of each location as the weight~\cite{Coutrot22}
\begin{equation} \label{eq6}
	{\mit \Gamma} = \frac{\sum_i O_i \gamma_i}{\sum_i O_i},
\end{equation}
where $\mit \Gamma$ is the centripetality of urban mobility, reflecting the overall orientation degree toward the city center of all flows. In contrast to previous studies~\cite{Louail15,Bassolas19}, our coarse-grained processing considers both the numerical and spatial information of mobility flows.

We further calculate the anisotropy and centripetality of mobility flows. Figs.~\ref{fig1}(d-f) show the results of three representative Chinese cities, namely, Beijing, Tianjin and Foshan, at a typical morning peak hour (hour 7). The home-work commuting mobility during the typical morning peak hour reflects the interaction between people’s residences and workplaces and contributes a large proportion of urban traffic. The PMVs in Beijing are relatively long, while the PMVs in Tianjin and Foshan are relatively short. This characteristic can be effectively quantified by the anisotropy of urban mobility: $\mit \Lambda_{Beijing}$=0.40, $\mit \Lambda_{Tianjin}$=0.28 and $\mit \Lambda_{Foshan}$=0.30. Furthermore, the PMVs in Beijing and Tianjin are more oriented toward the city center, while the PMVs in Foshan are relatively chaotic. This characteristic can be effectively quantified by the centripetality of urban mobility: $\mit \Gamma_{Beijing}$=0.92, $\mit \Gamma_{Tianjin}$=0.87 and $\mit \Gamma_{Foshan}$=0.65. These results suggest that $\mit \Lambda$ and $\mit \Gamma$ provide quantitative tools to objectively measure the overall characteristics of urban mobility. Thus, these two metrics together provide a coarse-grained representation of all mobility flows, which can reduce the complexity of detailed descriptions while preserving meaningful numerical and spatial properties of urban mobility systems.

\subsection{Spatial patterns of urban commuting flows}

The anisotropy and centripetality during the typical morning peak hour enable us to quantify the overall characteristics of commuting flows. First, each city is characterized by a point in the two-dimensional anisotropy-centripetality space (Fig.~\ref{fig2}a). We can measure the distance between these two points, which reflects the similarity of urban mobility characteristics~\cite{Louail15}. Then, we apply the hierarchical clustering method to cluster similar cities. Supplementary Fig. 4a shows a dendrogram resulting from the classification, where three well-separated clusters are identified and depicted in different colors in Fig.~\ref{fig2}a. We find statistically significant differences in anisotropy and centripetality across these three types of cities (Figs.~\ref{fig2}b and c), suggesting that the classification is qualitatively distinct. We additionally test the robustness of this classification against different hierarchical clustering approaches. The classification is not affected by the proximity measures used to combine clusters in hierarchical algorithms (see Supplementary Note 4, Supplementary Figs. 4b-c and Supplementary Fig. 5).

Furthermore, the classification has important implications for mobility patterns. Type 1 cities (upper right in Fig.~\ref{fig2}a) exhibit a high level of anisotropy and centripetality, indicating that the overall imbalance degree in the direction distribution and the overall orientation degree toward the city center are both strong. Such mobility behavior is commonly observed in cities with a monocentric configuration (e.g., Beijing, see Fig.~\ref{fig1}d), where there is a predominant central business district and people generally live in the suburbs~\cite{Louail14}, resulting in high commuting flows along radial directions toward the center. Hence, we refer to type 1 cites as having strong monocentric patterns~\cite{Thomson77}. Type 3 cities (lower left in Fig.~\ref{fig2}a) exhibit a relatively low level of anisotropy and centripetality, implying that the overall imbalance degree in the direction distribution and the overall orientation degree toward the city center are both weak. Such mobility behavior is commonly observed in cities with a polycentric configuration (e.g., Foshan, see Fig.~\ref{fig1}f), where the residential and workplaces are spatially mixed~\cite{Louail14}; therefore, commuting behavior is more isotropic and centrifugal. Hence, we refer to type 3 cites as having polycentric patterns. Type 2 cities (upper left in Fig.~\ref{fig2}a) appear to show a combination of low anisotropy and high centripetality, implying that the overall imbalance degree in the direction distribution is weak, while the overall orientation degree toward the city center is strong. Such mobility behavior is commonly observed in cities that are undergoing a transition from a monocentric to a polycentric configuration (e.g., Tianjin, see Fig.~\ref{fig1}e). In these cities, the development of suburbs leads to a relatively balanced distribution of residences and workplaces, resulting in more isotropic commuting behavior. However, the central business district still possesses many job opportunities and thus attracts a substantial inward flow of commuters. Hence, we refer to type 2 cites as having a weak monocentric pattern~\cite{Thomson77}. To further illustrate the differences in mobility characteristics, we visualize 6 representative cities for each city type (see Supplementary Note 5 and Supplementary Figs. 6-8). These visualizations show that cities within the same type share similar travel characteristics, while cities from different types exhibit differences in travel behavior.

\subsection{Commuting distance of different patterns}

To enhance the understanding of different mobility patterns, we explore the relation between the average commuting distance and city size. The average commuting distance is a crucial signature that is connected to not only the mobility efficiency but also the amount of gasoline consumed and CO$_2$ emitted by the city~\cite{Creutzig15}. Figs.~\ref{fig3}a-c depict scatter plots of the average commuting distance $D$ and urban area $A$ for the three types of cities. For two types of monocentric cities, the average commuting distance displays a significant increase with the size of the urban area (Figs.~\ref{fig3}a-b), with $R^2=0.84$, $p<0.001$ and $R^2=0.73$, $p<0.001$, respectively. This observation is an expected effect, as larger cities tend to have longer commuting distances~\cite{Louail15}, which is a form of urban allometric scaling known as large cities being scaled-up versions of smaller cities~\cite{Barthelemy19}. Most of these cities have expanded their urban area by gradually spreading out from the central business district. This growth pattern often leads to the expansion of residential areas, while most workplaces remain in their original locations~\cite{Louail15}. Therefore, the average commuting distance increases rapidly with city size under the combined effects of urban expansion and the separation of residences and workplaces. More interestingly, we note that the average commuting distance of weak monocentric cities increases slower than does that of strong monocentric cities. This indicates that when a city grows, individual people may have a greater increase in commuting distance in a strong monocentric pattern than in a weak monocentric pattern.

Another interesting observation is that there is no significance between the average commuting distance and the urban area for cities with a polycentric pattern, with $R^2=0.10$ and $p=0.35$ (Fig.~\ref{fig3}c). However, the average commuting distance actually increases initially and then remains approximately stable, which is similar to the relation between congestion-induced delays and city size~\cite{Depersin18}. This indicates a fundamental difference in growth and expansion patterns for polycentric cities and the other two patterns. Polycentric cities such as Foshan, Dongguan and Suzhou historically developed from the convergence of several villages, towns or counties. A merged city is more polycentric, with more mixed residences and workplaces; therefore, the average commuting distance remains relatively stable despite the increase in urban area. These trends can be confirmed by urban population size, another important indicator of city size. Figs.~\ref{fig3}d-f show that two types of cities with monocentric patterns still follow the allometry law, namely, the average commuting distance increases with increasing urban population. For cities with polycentric patterns, the average commuting distance remains relatively stable. These results indicate that our overall characteristics quantification method provides an appropriate classification of mobility patterns, with each pattern exhibiting different growth rates in the relationship between the average commuting distance and city size.

\subsection{Spatiotemporal variations in anisotropy and centripetality}

Having examined the overall characteristics of urban mobility at the city scale, we now investigate the spatial distribution patterns of the anisotropy and centripetality within the city at the mesoscopic scale. We first need to normalize the urban space to study cities of different sizes within a common frame of reference. Specifically, we define urban space as a hierarchy of $l$ levels~\cite{Aless20,Lee17} according to the principle of equal trip generation (see Fig.~\ref{fig4}a); that is, the sum of outflows originating from the cells included in each level is equal. Here, we use $l=5$ for all cities, but our results are robust against changes in this value (see Supplementary Note 6 and Supplementary Fig. 9). Then, we calculate the anisotropy and centripetality for each spatial level (Supplementary Note 7) and find some interesting phenomena.

\begin{figure*}
\centering
{\includegraphics[width=\linewidth]{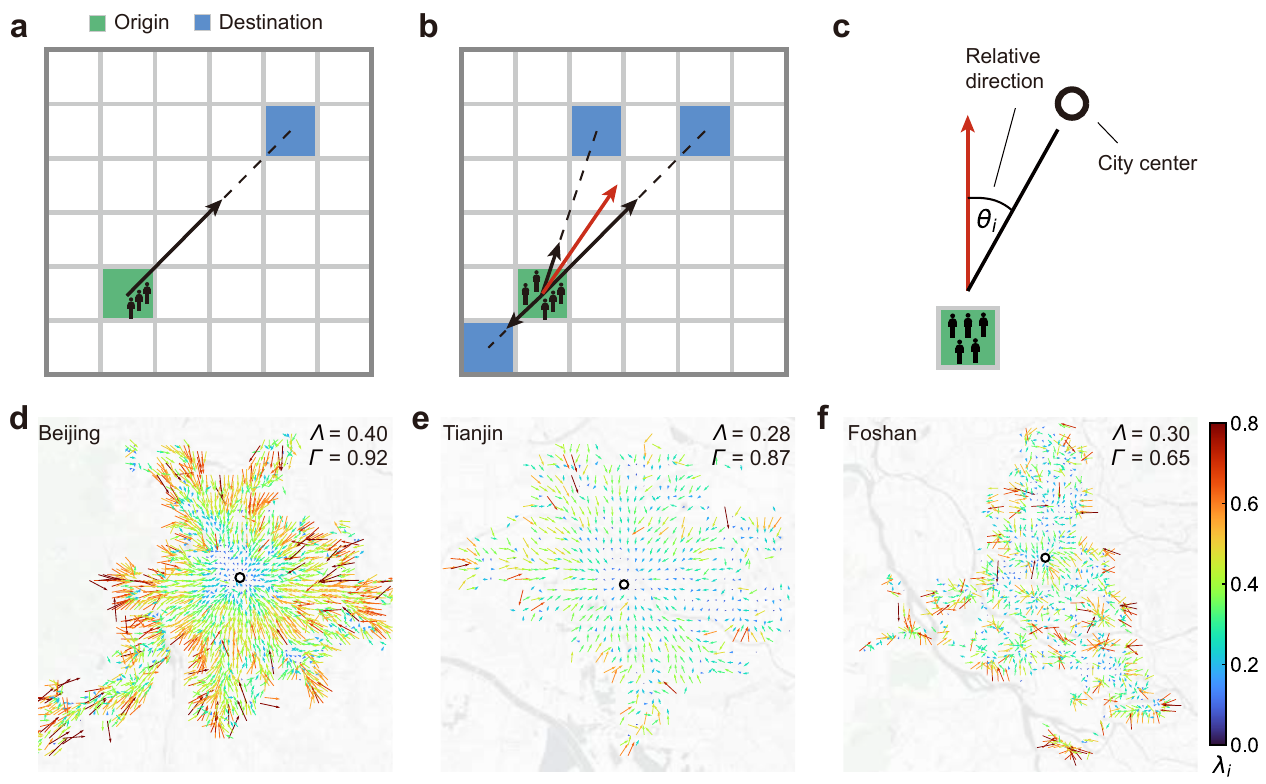}}
\caption{{\bf Population mobility vector.} 
	{\bf a} Schematic diagram showing the method to vectorize mobility flow. The direction of $\protect\overrightarrow{T_{ij}}$ (black arrow) is defined from the origin ($i$, green) to the destination ($j$, blue), and the magnitude is defined as the relative flow (i.e., the flow from $i$ to $j$ normalized by the outflow $O_i$)
	{\bf b} Definition of PMV. The origin $i$ has three destinations, and each flow from $i$ to $j$ can be represented as a vector using the method shown in {\bf (a)}. Taking a vector sum of these vectors, we obtain the PMV $\protect\overrightarrow{T_{i}}$ (red arrow).
	{\bf c} Definition of the relative direction. The relative direction ($\theta_i$) of the PMV is defined as the direction referenced to the city center (white dot), which can be calculated as the angle between the direction of the PMV (red arrow) and the direction to the city center (black line).
	{\bf d-f} Maps of PMVs for {\bf (d)} Beijing, {\bf (e)} Tianjin and {\bf (f)} Foshan. The length of the arrow is proportional to the magnitude of the PMV. The white dot represents the city center.
	}
\label{fig1}
\end{figure*}

\begin{figure*}
\centering
{\includegraphics[width= \linewidth]{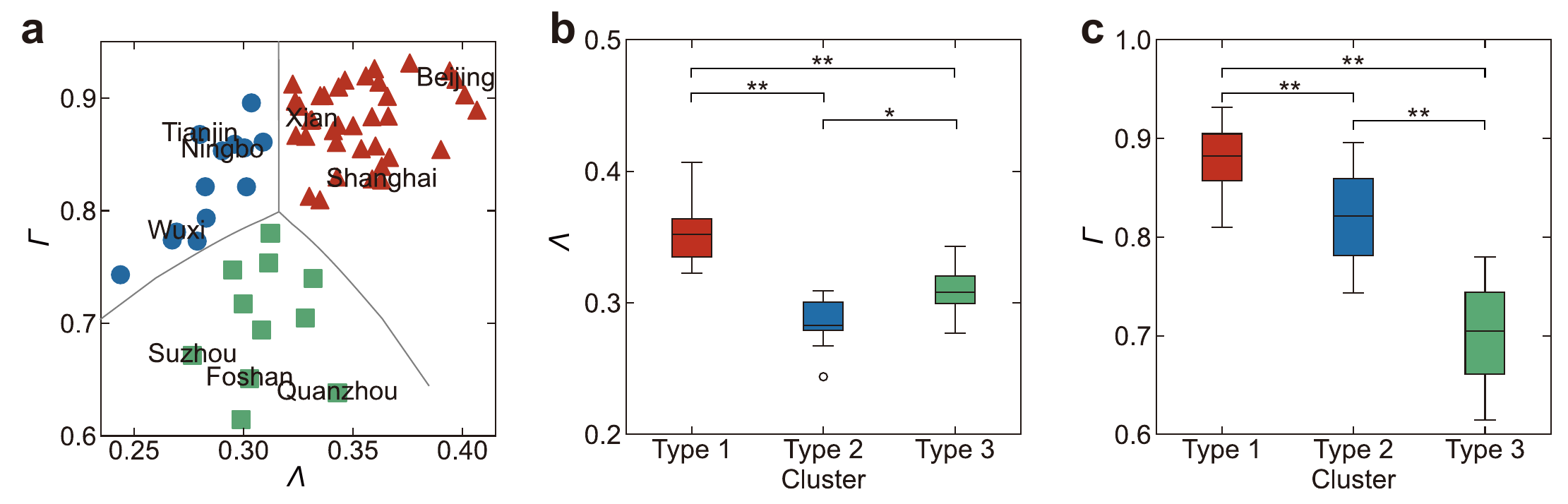}}
\caption{{\bf Hierarchical clustering of cities.}
	{\bf a} Classification of cities. The cities are divided into three types based on their anisotropy and centripetality of commuting flows. The colored symbols correspond to different types of cities: type 1 (strong monocentric, red triangles), type 2 (weak monocentric, blue circles) and type 3 (polycentric, green squares). The gray solid lines are shown as a guide to the eye.
	{\bf b} Comparison of anisotropy across type 1-3 cities. These three types of cities have statistically significant differences in anisotropy.
	{\bf c} Same as in {\bf (b)} but for centripetality.
	For the box plots in {\bf (b)} and {\bf (c)}, the central mark indicates the median, and the bottom and top edges of the box indicate the 25th and 75th percentiles, respectively; the whiskers extend to the most extreme data points within 1.5 times the interquartile range from the bottom or top of the box, and all more extreme points are plotted individually using a circular symbols. $^{*}P < 0.05$; $^{**}P < 0.001$.
	}
\label{fig2}
\end{figure*}

\begin{figure*}
\centering
{\includegraphics[width=\linewidth]{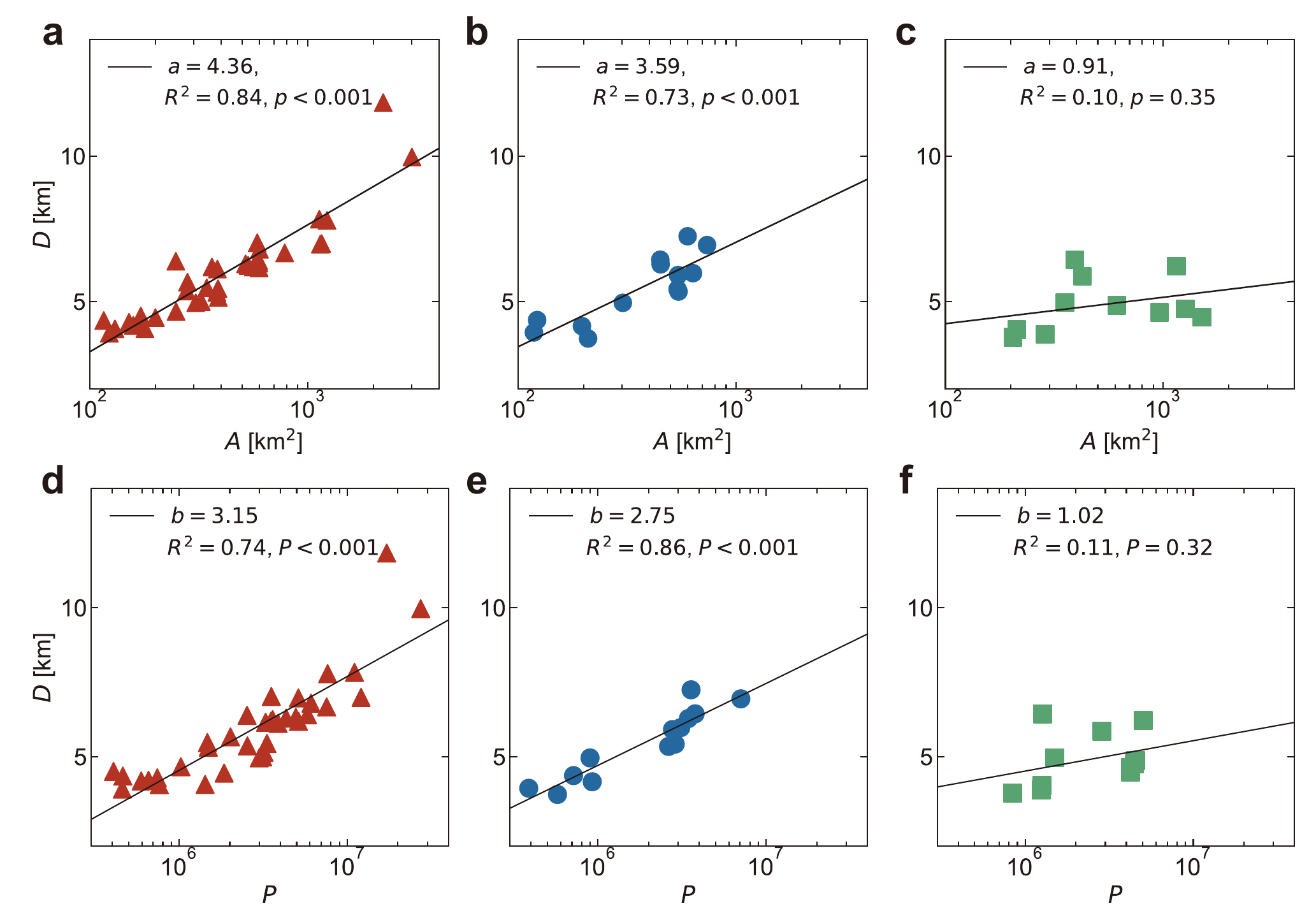}}
\caption{{\bf Average commuting distance and city size across city types.} 
	{\bf a-c} Average commuting distance versus area size for {\bf (a)} strong monocentric cities, {\bf (b)} weak monocentric cities and {\bf (c)} polycentric cities. The solid line indicates a logarithmic function fit ($D\sim \log A^{a}$) for which the parameter is provided in each panel.
	{\bf d-f} Average commuting distance versus population size for {\bf (d)} strong monocentric cities, {\bf (e)} weak monocentric cities and {\bf (f)} polycentric cities. The solid line indicates a logarithmic function fit ($D\sim \log P^{b}$) for which the parameter is provided in each panel. A linear regression {\it t} test was conducted to determine whether the slope of the regression line differed significantly from zero.
	}
\label{fig3}
\end{figure*}

\begin{figure*}
\centering
\includegraphics[width=\linewidth]{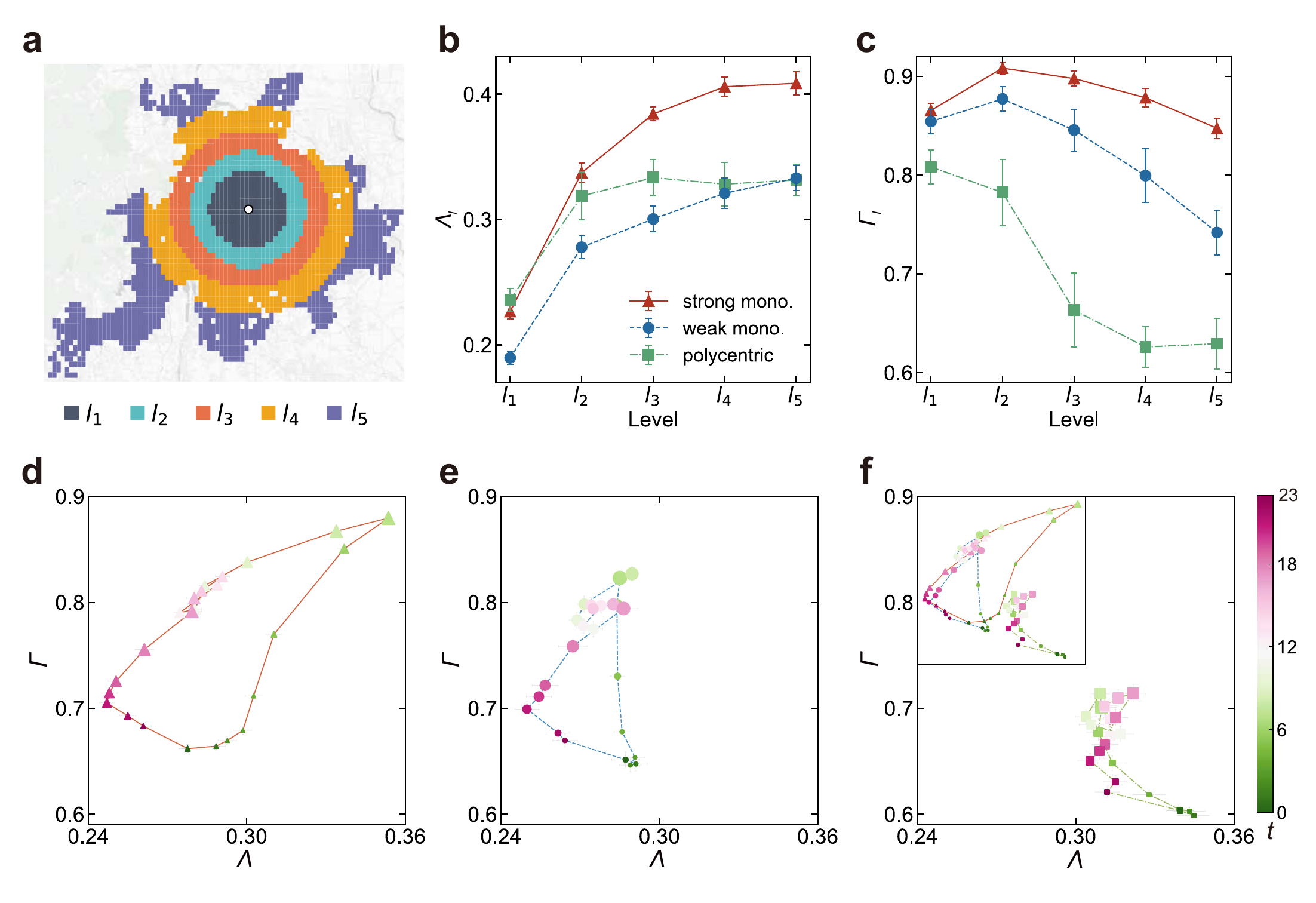}
\caption{{\bf Spatiotemporal variations in anisotropy and centripetality.}
	{\bf a} Illustration of the spatial hierarchical structure in Beijing. Each color corresponds to a space level, and trip generation of different levels is equal.
	{\bf b-c} The {\bf (b)} anisotropy and {\bf (c)} centripetality of each spatial level for the three types of cities. Error bars represent the standard error. Symbols and lines refer to various city types.
	{\bf d-f} Hourly anisotropy and centripetality of urban mobility for {\bf (d)} strong monocentric cities, {\bf (e)} weak monocentric cities and {\bf (f)} polycentric cities.
	Symbol sizes indicate the relative trip generation volume per hour. Symbol colors denote different hours.
}
\label{fig4}
\end{figure*}

First, the anisotropy systematically increases with the spatial level across city types (see Fig.~\ref{fig4}b and Supplementary Figs. 10-12), indicating that the imbalance degree in the direction distribution of commuting flows is relatively weak in the urban core, while such imbalance tends to become stronger in peri-urban areas. A plausible explanation for the observed trend is that job opportunities are not distributed evenly across space, with the urban core offering more balanced opportunities in each direction. Consequently, the anisotropy associated with human mobility is less pronounced in the urban core than it is in the urban periphery. In addition, polycentric cities remain relatively stable in anisotropy after level $l_2$, indicating that the direction distribution of job opportunities is approximately equal in most peri-urban areas of polycentric cities. Furthermore, strong monocentric cities experience the largest increase in anisotropy, with a growth rate of $80.3\%$, followed by weak monocentric cities ($75.5\%$), and finally polycentric cities ($40.5\%$). The varying degrees of increase in anisotropy indicate that the direction distribution of job opportunities in the peri-urban area of strong monocentric cities is the most imbalanced compared to their urban core (e.g., Beijing, see Fig.~\ref{fig1}d), followed by weak monocentric cities (e.g., Tianjin, see Fig.~\ref{fig1}e), and finally polycentric cities (e.g., Foshan, see Fig.~\ref{fig1}f).

Second, the centripetality systematically decreases with the spatial level across city types (see Fig.~\ref{fig4}c and Supplementary Figs. 13-15), indicating that the orientation degree toward the city center of commuting flows is relatively strong in the urban core, while such orientation tends to become weaker in peri-urban areas. This result points to the existence of a core-periphery structure in the city~\cite{Lee17}, where the attraction of the urban core to commuting flows gradually decreases from the city center to the periphery. In addition, two types of monocentric cities show relatively lower centripetality at level $l_1$, indicating that the dominant mobility within level $l_1$ exhibits weaker orientation toward the city center (on average). This is because these two types of cities have predominant central business districts within level $l_1$, which provide many job opportunities. Therefore, residents within level $l_1$ have more freedom of choice in their workplaces, resulting in a more diverse dominant mobility. Furthermore, strong monocentric cities experience the smallest decrease in centripetality with a decay rate of $2.1\%$, followed by weak monocentric cities ($13.2\%$), and finally polycentric cities ($22.1\%$). The varying degrees of decrease indicate that the urban core of strong monocentric cities is the most attractive (e.g., Beijing, see Fig.~\ref{fig1}d), followed by that of weak monocentric cities (e.g., Tianjin, see Fig.~\ref{fig1}e), and finally polycentric cities (e.g., Foshan, see Fig.~\ref{fig1}f).

Next, we explore the temporal variation patterns of anisotropy and centripetality associated with urban mobility for one day. The study scenario is still based on human mobility at the city scale. Figs.~\ref{fig4}d-f illustrate the average results of the hourly anisotropy and centripetality of urban mobility for the three types of cities (see Supplementary Figs. 16-18 for the details about individual city). The first striking observation is that strong monocentric cities display the largest variations in anisotropy and centripetality, followed by weak monocentric cities and finally polycentric cities. In particular, both types of monocentric cities show consistent patterns during off-peak hours (from 9 to 23 hours) but present large distinctions during morning peak hours (from 4 to 8 hours), and polycentric cities show pronounced differences compared to the other two types throughout the day.

Furthermore, despite the different values of anisotropy and centripetality, the trend is the same for the three types of cites during off-peak hours, suggesting that the temporal variation trend is an intrinsic and universal property of urban mobility during this period. For example, the anisotropy and centripetality show little variation from 9 to 17 hours, suggesting that the overall characteristics of urban mobility remain relatively stable during this period. Next, the anisotropy and centripetality weaken from 18 to 21 hours, primarily coinciding with the time when most people return home from work. This trend means that work-home flows become more isotropic and centrifugal during this period. Then, the centripetality is weakening while the anisotropy is increasing after 22 until early morning hours, mainly corresponding to the period when some people engage in nighttime leisure activities. This trend indicates that nighttime leisure flows become more anisotropic and centrifugal during this period.

During morning peak hours, the three types of cites exhibit considerable differences in the temporal variation patterns of anisotropy and centripetality. Although all three types of cites experience an increase in centripetality during this period, the degree of increase varies. Strong monocentric cities experience the greatest increase in centripetality, followed by weak monocentric cities, and finally polycentric cities. The varying degrees of increase in centripetality indicate that the urban core of strong monocentric cities is the most attractive, followed by that of weak monocentric cities and finally polycentric cities. On the other hand, regarding anisotropy, strong monocentric cities show an increase throughout this period, while that of weak monocentric cities remains relatively stable, and that of polycentric cities weakens. This result indicates that the dynamic state of the direction distribution of job opportunities in strong monocentric cities becomes increasingly imbalanced with the increase in trip volume during this period, while weak monocentric cities remain stable and polycentric cities become more balanced.

Furthermore, polycentric cities exhibit, on average, greater anisotropy than do weak monocentric cities throughout the day. This is due to the higher degree of directional consistency of human mobility within the core catchment areas of each subcenter in polycentric cities. For example, during the typical morning peak hour, the anisotropy is relatively large in Junan (a town of Foshan, see Supplementary Fig. 19). This can be attributed to the presence of several business centers, such as Junan cowtoy city, which is surrounded by numerous rural areas, including Xingcha village, Jitou village, and Beihai village. Residents living in these surrounding areas tend to commute to the business centers in town for work, leading to a higher degree of anisotropy for the town. Taken together, these spatiotemporal analysis results demonstrate that the classification of these three types of cities based on anisotropy and centripetality is appropriate.

\section{Model}

\begin{figure*}
\centering
{\includegraphics[width=\linewidth]{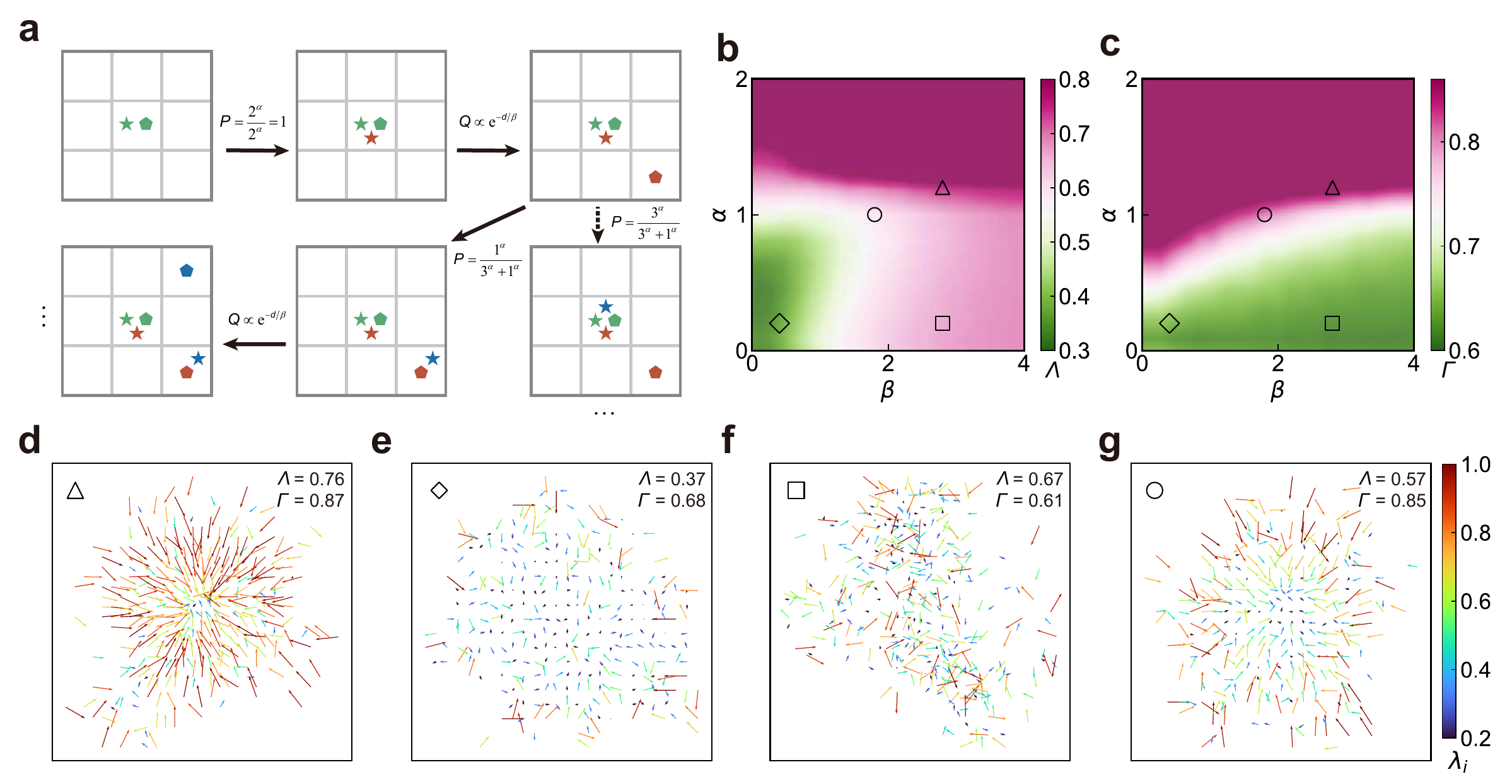}}
\caption{{\bf The random workplace and residence choice model.}
	{\bf a} Schematic representation of the RWRC model. Individuals are shown as different colored symbols. In the upper-left panel, the first (green) individual selects the central cell as the workplace (star) and residence (pentagon). Next, the second (red) individual selects a workplace according to the active population distribution. The probability that he or she selects the central cell is $P={2^{\alpha}} / {2^{\alpha}}=1$, as shown in the upper-middle panel. Then, the second individual selects a residence at distance $d_{ij}$ from his or her workplace with probability $Q$, as shown in the upper-right panel.
	Next, the third (blue) individual has a $P={3^{\alpha}} / (3^{\alpha} + 1^{\alpha})$ probability of choosing the central cell as a workplace, as shown in the lower-right panel, and a $P={1^{\alpha}} / (3^{\alpha} + 1^{\alpha})$ probability of choosing the lower-right cell as a workplace, as shown in the lower-middle panel. If the third individual selects the option in the lower-middle panel, he or she selects a residence at distance $d_{ij}$ from his or her workplace with probability $Q$, as shown in the lower-left panel.
	{\bf b-c} The values of {\bf (b)} anisotropy and {\bf (c)} centripetality obtained from the RWRC model with different parameters $\alpha$ and $\beta$. The color maps depict the mean value over 100 independent simulations. Each symbol represents a scenario with different values of $\alpha$ and $\beta$: ($\triangle$) $\alpha=1.2$ and $\beta=2.8$; ($\Diamond$) $\alpha=0.2$ and $\beta=0.4$; ($\Box$) $\alpha=0.2$ and $\beta=2.8$; ($\circ$) $\alpha=1$ and $\beta=1.8$.
	{\bf d-g} Maps of PMVs for these scenarios.
}
\label{fig5}
\end{figure*}

Our empirical analysis reveals that mobility flows exhibit different degrees of anisotropy and centripetality. In what follows, we build a random workplace and residence choice (RWRC) model to provide a mechanistic explanation for the diverse observations of the two metrics. Consider an $l \times l$ lattice in a 2-D Euclidean space, where $l$ is the number of locations in each dimension. Initially, the first individual chooses the center of the lattice as his or her workplace and residence~\cite{Li17,Xu21}. In each subsequent time step, a new individual is added to the simulation space. The newly added individual first chooses his or her workplace according to employment opportunities~\cite{Simini12} and then chooses his or her residence according to commuting costs~\cite{Louf13,Wang21}. We assume that each new individual makes these two choices (see Fig.~\ref{fig5}a). (\romannumeral1) Choosing workplace: the individual chooses location $j$ as the workplace with probability
\begin{equation} \label{eq7}
	P_{j} \propto {N_{j}}^{\alpha},
\end{equation}
where $N_j$ is the sum of the residential and working populations of location $j$, which is also called the active population and is an appropriate proxy for estimating socioeconomic opportunities~\cite{Li17}. $\alpha \geq 0$ is a parameter characterizing the strength of population attraction for employment. This workplace selection rule is an assumption known as preferential attachment in network and social science, which embodies the intuitive idea of “the rich get richer” dynamics~\cite{Song10,Bara99}. (\romannumeral2) Choosing residence: the individual chooses location $i$ ($i\neq j$) as the residence with probability
\begin{equation} \label{eq8}
	Q_{ij} \propto \mathrm{e}^{- d_{ij} / \beta},
\end{equation}
where $d_{ij}$ is the distance between $i$ and $j$ and $\beta > 0$ is a typical scale parameter characterizing commuting distance~\cite{Barthelemy11}. This residence selection rule is reasonable, as the commuting cost naturally increases with distance, implying that the probability of an individual working at $j$ choosing to live at $i$ decreases as the distance $d_{ij}$ increases. Here, we use exponential decay with a finite typical scale~\cite{Barrat05}.

We perform numerical simulations for our model in a system with a population of ${10}^4$ individuals on a regular grid of $50\times50$ cells. For employment strength $\alpha$, we select a typical range of values that span from sublinear to superlinear scalings, covering the spectrum from 0 to 2 with a step size of 0.1. For distance scale $\beta$, we start at 0.01 and increase the value by a step size of 0.1 until it exceeds 4, after which the anisotropy and centripetality are approximately stable. Figs.~\ref{fig5}b-c show the values of anisotropy and centripetality obtained when changing model parameters $\alpha$ and $\beta$. When employment strength $\alpha$ is relatively large (triangle in Fig. 5b-c), the majority of added individuals tend to choose the central areas of the simulation space as their workplaces. As a result, commuting flows exhibit strong anisotropy and centripetality. An example with high values of $\alpha$ and $\beta$ is shown in Fig.~\ref{fig5}d, which is consistent with the strong monocentric pattern in empirical observations (see the example in Fig.~\ref{fig1}d). When both employment strength $\alpha$ and distance scale $\beta$ are relatively small (diamond in Fig. 5b-c), the majority of added individuals tend to choose a workplace uniformly among different areas and reside near their workplace. As a result, compared to the triangle scenario, commuting flows exhibit weaker anisotropy and centripetality, as seen in the example in Fig.~\ref{fig5}e. When employment strength $\alpha$ is relatively small and distance scale $\beta$ is relatively large (square in Fig. 5b-c), residential locations that provide job attraction are more dispersed throughout the entire space. As a result, compared with the diamond scenario, the centripetality is weaker, while the anisotropy increases, as seen in the example in Fig.~\ref{fig5}f, which is consistent with the polycentric pattern in empirical observations (see the example in Fig.~\ref{fig1}f). When employment strength $\alpha$ and distance scale $\beta$ are both moderate (circle in Fig. 5b-c), commuting mobility exhibits a combination of low anisotropy and high centripetality, as seen in the example in Fig.~\ref{fig5}g, which is consistent with the weak monocentric pattern observed in the empirical observations (see the example in Fig.~\ref{fig1}e). These findings show that the RWRC model can successfully reproduce the observed overall commuting characteristics, suggesting that the strength of population attraction for employment and the scale of commuting distance are two fundamental principles responsible for shaping commuting patterns.

\section{Discussion}

The past decade has witnessed considerable effort to quantify and understand the overall characteristics of urban mobility~\cite{Louail15,Louail14,Lee17,Bassolas19,Shida20,Shida22,Coutrot22,Mazzoli19,Hame19}. Previous quantification efforts have mostly aimed to extract coarse-grained numerical information from the OD matrix to characterize urban mobility without incorporating the spatiality of mobility~\cite{Louail15,Bassolas19}. However, spatial and numerical information is closely correlated~\cite{Hevia14}. This ignorance of spatial or numerical information may lead to misconceptions about the overall characteristics of human mobility and thus may have far-reaching practical consequences for urban planning, traffic management and many other applications~\cite{Schlapfer21,Wieder16}. For example, recent evidence suggests that the spatial layout of a system has a significant effect on the system’s structure and function~\cite{Liu21}. In particular, mobility networks with the same OD matrix but different spatial layouts tend to exhibit distinct overall characteristics. The present paper seeks to address this critical need for formulating a coarse-grained method that can incorporate both spatial and numerical information associated with human mobility. We use vectors to describe human mobility, where the direction of mobility reflects spatial information and the size of mobility reflects numerical information. As a result, we reduce the OD matrix into two metrics, namely, anisotropy and centripetality, enabling the quantification of the overall imbalance in direction distribution and orientation toward the city center of mobility flows. Together, these two metrics provide a comprehensive representation of urban mobility and offer important insights into urban rhythm~\cite{Louail14}.

The anisotropy and centripetality of urban mobility enable us to systematically compare different mobility characteristics. We applied these two metrics to quantify the overall characteristics of mobility flows in 60 major Chinese cities. We clustered the mobility patterns of these cities into three types, i.e., strong monocentric, weak monocentric and polycentric cities, according to the anisotropy and centripetality during the typical morning peak hour. Further, we studied the relationship between the average commuting distance and city size across different types of cities. We found that the average commuting distance of strong monocentric and weak monocentric cities displays a significant increase with city size and follows an allometric scaling relationship, whereas that of polycentric cities remains relatively stable, indicating that the mobility efficiency of polycentric cities is higher than that of the other two types of cities. This result supports the belief that polycentric evolution has positive effects on commuting efficiency~\cite{Jun20}. Furthermore, we presented a thorough investigation of the spatiotemporal variations in anisotropy and centripetality, which provides a detailed picture of urban mobility and deepens our understanding of various mobility patterns. Finally, we built a microscopic model that can reproduce the diverse empirical observations of anisotropy and centripetality, establishing a link between workplace-residence choice behavior at the individual level and macroscopic characteristics of urban mobility. These results could help to advance our understanding of the structure and evolution of urban mobility systems, which would have potential applications for urban and transportation planning.

It is important to note that in contrast with the hotspot flow matrix~\cite{Louail15} and flow-hierarchy~\cite{Bassolas19}, our method directly considers the spatial information of mobility. The hotspot flow matrix and flow-hierarchy both quantify the organization of urban mobility by calculating the flow proportion between hotspots of varying levels of activity. These metrics uncover unique structural characteristics of urban mobility, but they do not directly consider the spatial information associated with human mobility. Fortunately, inspired by the route geometric metric, which encapsulates the direction, orientation and length of travel routes~\cite{Lee17}, we proposed a PMV that can capture the direction, orientation (relative to the city center) and extent of mobility. Interestingly, the PMV is the same as the vector field for recurrent mobility~\cite{Mazzoli19}. However, although we use the same mathematical form as the vector field, the research questions that we investigate are different. The vector field is designed to verify whether commuting mobility fulfills Gauss’s theorem and is irrotational, and after confirming these factors, a scalar potential is defined to determine separate mobility basins and discern contiguous urban areas. By contrast, our aim is to define anisotropy and centripetality metrics according to the PMV to help quantify the overall characteristics of mobility flows. We believe that the present method represents an important step toward providing quantitative and comprehensive insight into urban mobility patterns, which is significant for sustainable development goals.

Our results have significant consequences for both practical and theoretical applications. Given the lack of consensus in rigorously measuring the organization of cities, a wealth of studies propose various metrics from multiple dimensions, such as population~\cite{Volpati18}, infrastructure~\cite{Zhou22} and trip flows~\cite{Bassolas19}. The concepts of anisotropy and centripetality introduce a different approach based on trip flows and can help to reveal new insights into the organization of cities. In addition, our metrics can serve as fundamental tools to address hard open problems in urban science, such as the definition of the city center~\cite{Mazzoli19}. Here, we introduce a data-driven method and view the city center as the location where the average mobility direction of all flows is minimized (Methods). Our method offers an automated and systematic means of identifying the city center based on human mobility, thereby eliminating the need for subjective definitions or relying on coordinates provided by mapping services~\cite{Lee17,Coutrot22}.

There are numerous possible directions for future work on this topic. For example, we believe that an important feature of the quantification method is its versatility; hence, we can utilize it to study not only urban human mobility but also urban freight trips~\cite{Yang22}. This approach could provide valuable insights into the demand for urban freight and aid in the development of more effective policies for managing the transportation of goods within cities. Furthermore, the proposed RWRC model is built upon simple workplace-residence choice rules to account for the key mechanisms driving the diversity in anisotropy and centripetality. However, it is unable to explain the more intricate details of observed phenomena, such as scaling laws and temporal variation patterns. These characteristics of urban mobility require more advanced and sophisticated models that account for various relevant factors, such as individual preferences, socioeconomic factors, and temporal dynamics. By incorporating these factors, more accurate and comprehensive models can be developed to better explain and predict the observed phenomena in urban mobility.

\section{Methods}

\subsection{Data description}\label{methods1}

The aggregated and anonymized mobile phone dataset was provided by a Chinese telecommunications operator. The operator partitioned all cities into a $0.005^{\circ}$ ($\sim0.5~km\times0.5~km$) grid cell. The aggregated, nonpersonally identifiable human movement records represent the number of trips between different grids per hour in August and November 2019 in 60 Chinese cities (Supplementary Table 1). Each record contains the date, the hour, the coordinates (longitude and latitude) of the origin grid, the coordinates of the destination grid, and the number of trips. To protect customers’ privacy, no individual information or records are available in this dataset.

\subsection{Data preprocessing}\label{methods2}

To obtain the characteristics of human mobility, we process the dataset from two dimensions of space and time. In the spatial dimension, we merge the grid into a size of $\sim1~km\times1~km$. In the temporal dimension, from the mobile phone dataset obtained for 25 typical weekdays (Tuesday through Thursday), we calculate the average hourly trips between grids. In addition, human mobility follows the urban rhythm with high numbers of trips during the day and low numbers during the night. To illustrate our method, we thus select the time span between 7:00 and 7:59, which corresponds to the typical morning peak hour, to quantify the overall characteristics of urban mobility and to classify cities.

\subsection{Extracting the urban area}\label{methods3}

The mobile phone dataset used for this study was recorded according to the administrative region of each city. However, these administrative regions are not subject to either socioeconomical or morphological factors. To obtain a unified and harmonized delineation of cities, it is necessary to extract urban areas that go beyond the administrative regions defined by national authorities. Here, we adopt the global human settlement dataset, which defines the urban area as a contiguous area with a density of at least 1500 inhabitants per square kilometer or a majority of built-up land cover coincident with a minimum of 50000 inhabitants, to extract the urban area~\cite{Pesaresi16}. Supplementary Note 2 and Supplementary Fig. 2 provides a detailed description of the extracted results.

\subsection{Identification of the city center}\label{methods4}

Here, we develop a data-driven method, called the minimum mobility direction algorithm (MMDA), to identify the city center. We define city center $C$ as the location where the average mobility direction of all flows in the urban area is minimized. The algorithm procedure is as follows.

Step 1: Select an initial center $i_c$ and calculate the average mobility direction of flows originating from other locations to $i_c$, i.e., ${\sum_{i\neq i_c}O_i\theta_i} / \sum_{i\neq i_c}O_i$.

Step 2: For each location in the city, repeat step 1.

Step 3: Among all locations, select the location corresponding to the smallest average mobility direction as the city center.

Our algorithm provides an automated and systematic means of identifying the city center using human mobility data. The results identified with the MMDA are shown in Supplementary Fig. 3 and Supplementary Table 1. The outflows of the surrounding locations of the city center are relatively large, indicating that the identification result is reasonable.

\section*{Acknowledgments}

HW is supported by the National Natural Science Foundation of China (72271019, 41925003).
PZ is supported by the National Natural Science Foundation of China (41925003, 42130402) and Shenzhen Science and Technology Innovation Commission (JCYJ20220818100810024).
XYY is supported by the National Natural Science Foundation of China (72271019).

\section*{Author contributions}
H.W., P.Z. and X.-Y.Y. designed the research; H.W. and X.-Y.Y. performed the research; H.W., P.Z. and X.-Y.Y. contributed analytic tools; H.W., P.Z. and X.-Y.Y. analyzed data; and H.W., P.Z. and X.-Y.Y. wrote the paper.

\section*{Additional information}
\noindent
{\bf Competing financial interests:} The authors declare no competing financial interests.

\end{document}